# Next-Generation Industrial Networks: Integrating Time-Sensitive Networking for Smart Factory Reliability

**Yazen S. Sheet, Mohammed Younis Thanoun, Firas S. Alsharbaty**

Department of Electrical Engineering, University of Mosul, Mosul, Iraq

## Abstract:

The traditional industrial communication networks may not meet the requirements of the main smart factory applications together, such as Remote control and safety applications (which considered as strict real time applications) and augmented reality (which consumes wide bandwidth). This work suggests an enhanced communication network to serve an optimum case for the smart factory includes heavy data applications and real-time applications in one hand using the concepts of time sensitive networks (TSN) to address the limitation of real-time applications and the edge computing to assimilate the heavy data applications. The current work submits an experimental scenario to exploit H.265 compression method based on edge computing concept to mitigate consuming the capacity on overall network performance for augmented reality application. The results of enhanced communication network indicated that the latency of real time applications was less than 1msec while the packet data delivery of rest applications was 99.999%.

## 1- Introduction:

A Smart factory which considered as one of Industry 4.0 revolution concepts represents a groundbreaking shift in manufacturing sector, characterized by the integration of digital technologies and automation to create highly efficient, flexible, and intelligent production environments. It utilizes IIoT devices to connect machinery, sensors, and systems, enabling real, semi real and non-real time data exchange, this connectivity allows for seamless communication across the production line, facilitating quick responses to changes and challenges[1]. There are number of applications in smart factory that require different requirements in terms of delay, bandwidth and reliability of data transmission. Some of these applications require bounded latency requirements, for instance, a robotic arm which should pick up an object that moves on a conveyor belt at a specific time, while other may require high bandwidth and reliability, such as the AR application. Table -1- shows the most essential applications of the smart factory and its requirements in terms of delay, reliability, and data rate[2-4]. To address the requirements of these applications, it is necessary to handle a reliable communication network that is capable of delivering such data within a specific time that has been defined, which is considered a major challenge facing researchers in this domain. Fieldbus communications network like Controller Area Network (CAN), and Industrial ethernet like EtherCat, have been developed in the previous time in order to accomplish this objective[5, 6]. The main problem of such protocols is missing the compatibility between them; therefore, a new approach has been developed in 2012 by a group of researchers working inside the IEEE Working Group, this approach is called Time Sensitive Network (TSN) with standard IEEE 802.1. Through the context of this concept, the data is separated into several categories based on the kind of application being used, and it is then kept in distinct queues according to the relative significance of each category. These types are served by a variety of mechanisms and methods, which include mechanisms to transmit data from queues, such as the Strict Priority (SP) as specified in standard 802.1Q and Credited Based Shaping (CBS) algorithm that defined in standard 802.1av, as well as mechanisms that deal with queues itself, such as Time Aware Shaping (TAS) with standard 802.1bv and Frame Preemption as introduced in standard 802.1bu. This trend contributes in transmit different kinds of sensitive and non-sensitive applications on a same network, while at the same time maintaining the speed of traditional ethernet networks[7-9].

**Table -1- The Requirements of main Smart factory Applications**

| Application | Requirements | | | | |
|---|---|---|---|---|---|
| | Latency | Bandwidth | Availability | Mobility | Density |
| Automated Guided Vehicles | Medium (10-20ms) | Variable (<100Mbps) | 99.9999% | Needed | Low |
| Augmented Reality | Medium (10-20ms) | High (<1Gbps) | 99.99% | Needed | Depending on No. of Workers |
| Remote Control | High (0.1-1ms) | Low | 99.9999% | Not Necessary | Medium |
| Monitoring and Predictive Maintenance | Low (20-100ms) | Variable | 99.99% | Not relevant | High |
| Safety and Protection | Medium (1ms) | Low (<1Mbps) | 99.9999% | Not Necessary | Medium |

Consequently, many authors have examined these mechanisms in their research papers, implemented them, and evaluated their performance in different industrial scenarios. For example, the papers[10, 11] presented a comprehensive framework for the integration of Time-Sensitive Networking (TSN) features into standard Ethernet, emphasizing its potential to satisfy the QoS demands of industrial applications. It provides simulation experiments that demonstrate the efficacy of TSN features, including strict priority (SP), frame preemption (FP) and time-aware shaping (TAS), the results showed that TAS typically outperformed FP and ST regarding latency and jitter, while it imposed considerable configuration overhead. The authors in [12-14] designed an extensive simulation model utilizing OMNET++ and MATLAB simulation packages, specifically intended to assess Time-sensitive Networking protocols and their efficacy in industrial applications. They presented the implementation of GCLs of the IEEE 802.1 Qbv standard in the simulation model as well as frame preemption (FP) to guarantee deterministic transmission of time-sensitive data. The authors in [15-17] offered a complete analysis of several Time-Sensitive Networking (TSN) latency management methods, such as Frame Preemption, Credit-Based Shaper (CBS), and Gate management List (GCL). They examined the influence of these systems on end-to-end (E2E) latency. They presented optimization methods employing a surrogate model to ascertain optimal configurations for TSN networks, hence enhancing end-to-end latency to address the varied Quality of Service (QoS) demands of many data streams produced during the quality inspection process in smart factory. The works performed comprehensive simulations across multiple scenarios to assess the performance of many network configurations. The paper[18] examined the performance of TSN networks in IIoT environments by utilizing common topologies and message protocols, suggesting scenarios to evaluate traffic management. It offered a comparative review of various technologies employed for modeling services, emphasizing latency, channel efficiency, jitter, and queuing behavior across different congestion levels. The article examined the application of Edge Computing for improving network performance, presenting scenarios for both centralized and edge deployment strategies to improve service delivery. The works mentioned earlier have shortcomings in that they do not utilize a complex communication network capable of supporting the main applications of the smart factory in terms of reliability and latency requirements of these applications, as they concentrated solely on supporting time-sensitive applications within a simple network model disregarding the other applications and their effect on real-time applications, as well as the deficiencies of establishing time-sensitive network mechanisms and analyzing their performance in detail. However, the contributions of this research paper in comparison with these works can be summarized as follow:

- This work offers a complex communication network that deals with the essential candidate applications for smart factory in the era of industry 4.0.
- The current work handles the effect of non-real time applications on real time applications in terms of the applications requirements.
- The concept of TSN is captured by the current work to support the performance of industrial communication network.
- To assimilate the heavier augmented reality application (AR) in the industrial network, this research paper employs the emerging technologies in terms of edge computing and data compression to behave the performance of industrial communication network to meet the requirements of such applications.
- Utilizing the H265 video compression technology, which is based on the concept of edge computing in a practical manner on AR video data which has been taken in an industrial environment to calculate the appropriate compression ratio without affecting the quality of the augmented reality data, which has been also calculated practically to achieve a balance between the compression ratio and the quality of service for transmission over a communications network.

Table-2- illustrates the main contributions of related works compared to this work.

The structure of this paper is divided into six sections, besides an introduction. A description of the TSN and emerging technology have been given in Section 2 and 3 respectively, while section 4 has provided an explanation of the research methodology, section 5 presents the adopted scenarios, section 6 discusses the results. Lastly, section 7 explains the most significant conclusions of the work.

**Table -2- Comparison among this work and previous related works**

| Ref. | Adopted Algorithms | | | | | Emerging-Technology | Applications Kind | | | |
|---|---|---|---|---|---|---|---|---|---|---|
| | | TSN Mechanisms | | | | | | | | |
| | (FIFO) | SP | CBS | TAS | FP | | RT | SRT | NRT | BE |
| [10] | √ | √ | | √ | √ | | √ | | | √ |
| [11] | | √ | | √ | | | √ | | | √ |
| [12] | | | | √ | | | √ | | | √ |
| [13] | | √ | | √ | √ | | √ | | | √ |
| [14] | | | | √ | √ | | √ | | | √ |
| [15] | | √ | √ | √ | √ | √ | √ | √ | | √ |
| [16] | | | | √ | √ | | √ | √ | | √ |
| [17] | √ | √ | √ | √ | | | √ | √ | | √ |
| [18] | √ | | √ | | | √ | √ | √ | √ | √ |
| This Work | √ | √ | √ | √ | √ | √ | √ | √ | √ | √ |

FIFO: First In First Out, RT: Real Time, SRT: Semi Real Time, NRT: Non-Real Time, BE: Best Effort

## 2- Time-Sensitive Networks (TSN)

Time sensitive network is a communication network that carries out a time-bound transfer of information using the standard ethernet technology. TSN has been developed by researching group within IEEE working group under IEEE 802.1 Standard. This technology is used in industrial applications, aircraft, vehicles, health and transport fields. It operates within the second layer (MAC Layer) of OSI Layers Ref.) in which a 4 bytes tag has been added to standard Ethernet header. TSN have different standards to meet its functionalities[7]. The IEEE 802.1AS generalized precision time protocol (gPTP) is a standard which has been derived from IEEE 1588 that is more efficiently designed for applications that require precise timing. Similar to 1588, the 802.1AS protocol manages synchronization among a sequence of nodes that are linked by ethernet cables, switches, and bridges [9]. According to IEEE 802.1Q, standard bridging uses a strict priority (SP) algorithm as a default transmission selection approach for eight distinct traffic classes, each allocated a unique priority level. Frames belonging to the same traffic class are often selected using a First-In-First-Out (FIFO) approach. The priority levels are established based on the value of the Priority Code Point (PCP) field in the 802.1Q tag of a new ethernet frame, eight types of services can be defined due to this field Table -2- represents the mapping between priority and traffic classes. The original Ethernet frame header is appended with this tag to make TSN frame header[19]. Figure 1 explains that.

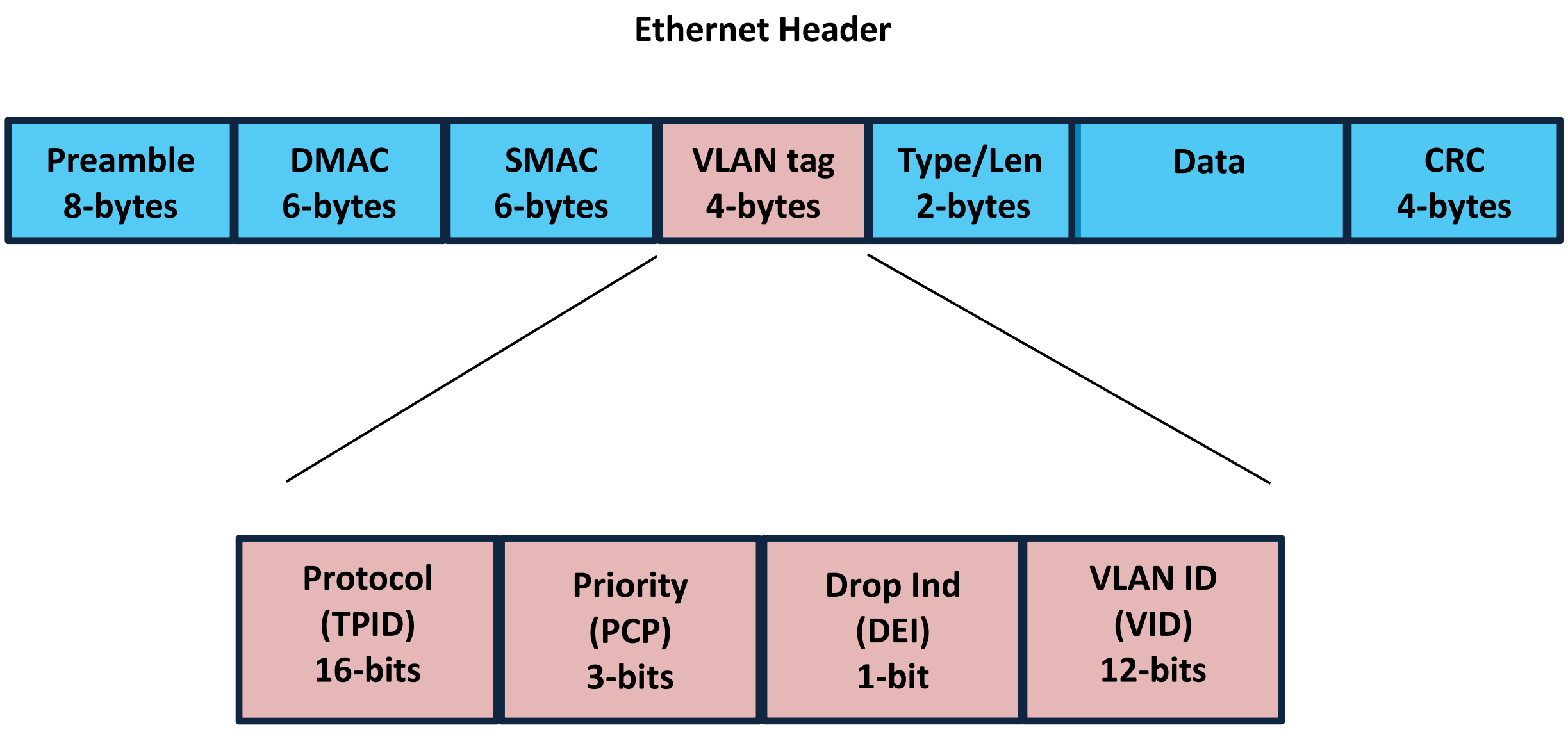

**Fig -1- IEEE 802.1 Q Tag**

**Table-3- Priority and Traffic class Map**

| Traffic Type | PCP | Priority |
|---|---|---|
| Background | 1(lowest) | 0 |
| Best effort | 0(default) | 1 |
| Excellent effort | 2 | 2 |
| Critical Application | 3 | 3 |
| Video | 4 | 4 |
| Voice | 5 | 5 |
| Internetwork Control | 6 | 6 |
| Network Control | 7 (highest) | 7 |

According to this algorithm, the highest priority queue frames have been selected for transmission over the rest of the queues with lower priority [20]. Another algorithm for frame transmission, called Credit-Based Shaper (CBS) under IEEE 802.1Qav standard, the transmission process as this algorithm does not depend only on the value of the queue's priority, but rather than on what is called credit. This algorithm added a kind of justice to other queues with lowest priority by dedicating a specific capacity for each queue that uses this algorithm. According to this algorithm, there are two significant terms [21]:

**IdleSlope:** which computes the rate at which a flow accumulates credits. It occurs during the period while frames are still awaiting transmission in a queue, IdleSlope is crucial in deciding the amount of bandwidth allocated to a flow. Bandwidth fraction for a flow can be calculated as eq (1):

$$\text{bandwidthFraction} = \text{IdleSlope}/\text{portTransmitRate} \text{ ------ } (1)$$

**SendSlope:** It represents the rate of credit decrease when frames are being transmitted from the queue and can be calculated by Equation (2):

$$\text{SendSlope} = \text{idleSlope} - \text{port_transmission_rate} \text{--------}(2)$$

Time aware shaper (TAS) or IEEE 802.1Qbv standard is regarded as the most important standard in TSN which has received significant attention from researchers. This standard has been utilized for supporting time-sensitive scheduled traffic, such as remote-control applications in many domains including factories, autos, transportation and etc. According to this standard, there is a gate for each of eight queues, each gate has two possible states: open or closed. The operation of these gates is controlled by a specialized module known as the Gate Control List (GCL). When the queue gate is in a closed state, sending from this queue is not possible. On the other hand, when the gate is open, transmission occurs from the queue associated with this gate, which ensures that time-critical messages will reach their destination within a specific amount of time. There are three periods during one cycle of sensitive applications transmission: Protected window, Unprotected window and Guard band [22]. Figure 2 illustrates the concept of this standard.

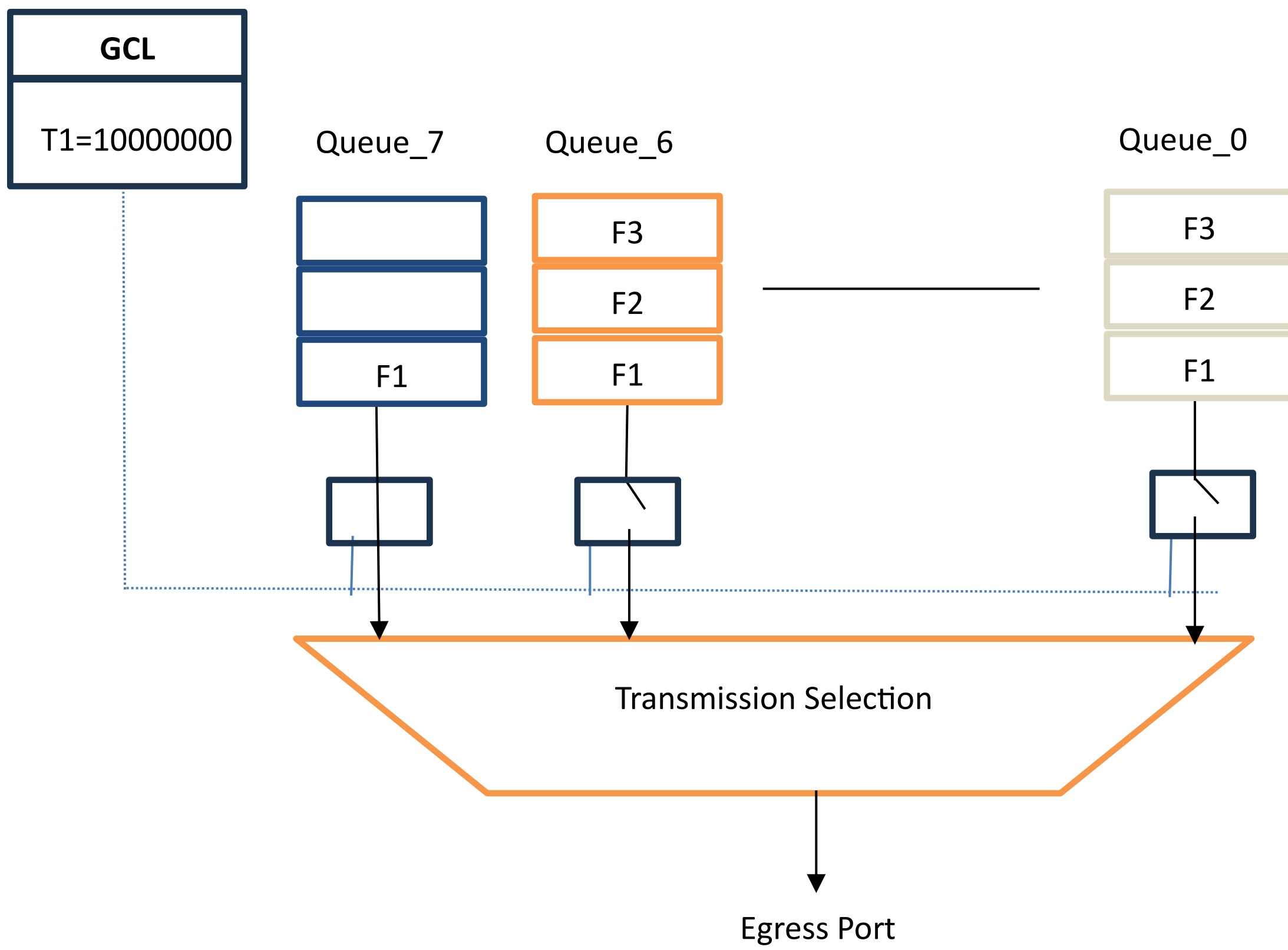

**Fig -2- TAS Concept**

Frame Preemption is another TSN standard that has been identified according to IEEE 802.1Qbu which employed to decrease the time delay for frames associated with time-sensitive applications of higher priority by interrupting and halting the transmission of frames for applications of lower priority upon the arrival of a higher-priority frame. Transmission of these lower frames resumes once the higher-priority frames have been sent, this operation is iterated multiple times whenever a frame with a higher priority has been arrived while transmitting frames with a lower priority [23].

## 3- Emerging Technologies in Indusrty 4.0

Emerging technologies in Industry 4.0 are revolutionizing manufacturing and industrial processes through the integration of advanced digital technologies like IIoT , AI and ML, Big data analytic , Edge computing and so on. Edge computing is one of these technologies that has been used in this work, the philosophy of edge computing focuses on executing computations nearby the data source. Edge computing resources may consist of a network or computer resources operated between end users and cloud data centers. IIoT devices generate enormous amounts of raw data that may exceed the capacity of the traditional cloud computing paradigm, this means that the majority of IIoT data will be consumed at the network edge rather than being transferred to the cloud. On the other hand, the cloud computing model may be insufficient for many reasons. For instance, the volume of data at the edge is excessive, using significant bandwidth and computational resources unnecessary. Moreover, the necessity for privacy protection obstructs cloud computing in IIoT. Therefore, the edge computing concept have been emerged to handle the mentioned cases[24]. Figure 3 illustrates the concept of Edge Computing.

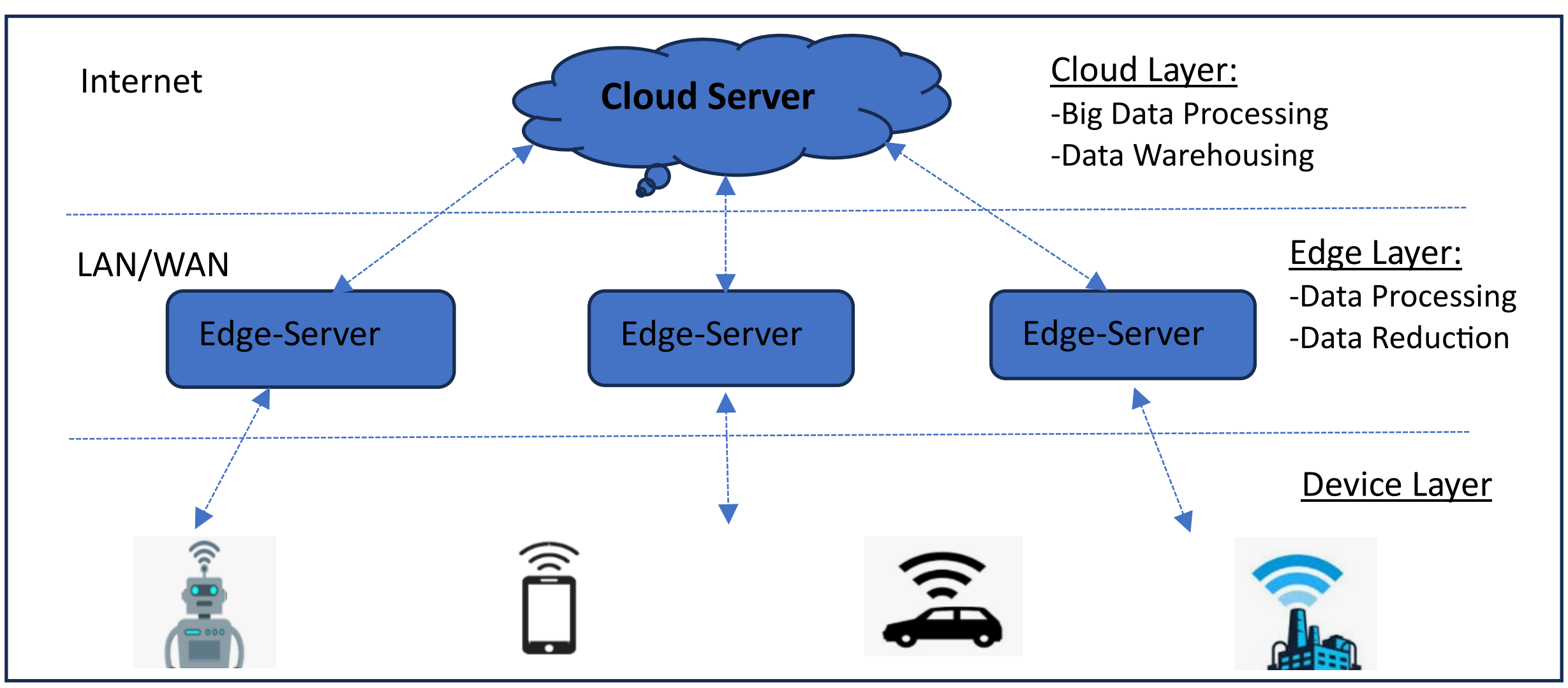

**Fig -3- Edge Computing Concept**

In this context, data compression is one approach that could be used based on edge computing concept to enhance system performance. The most prevalent form of data compression is video compression, which uses one of several regularly standards, H.264/AVC, VP9, AV1, or H.265/HEVC [25]. Using H.265/HEVC, streaming services can improve user experience and reduce data transmission costs by delivering higher-quality videos with manageable bandwidths. H.265 is the codec that would replace H.264, where its compression ratio is up to 50% higher than H.264, making it useful in a lot of different places. Compression solutions that rely on edge computing do have some downsides, though one of these is the processing time required for encoding and compression, which is around 20 milliseconds per megapixel when using a 50:1 compression ratio[26].

## 4- Methods and Materials

### 4.1- Model Description

The smart factory communication network that has implemented in this work addresses an industrial unit that is made up of a collection of components that are representative of the most important applications of a smart factory. The structure of the various components of the designed model and the flow of messages has been presented in Figure 4.

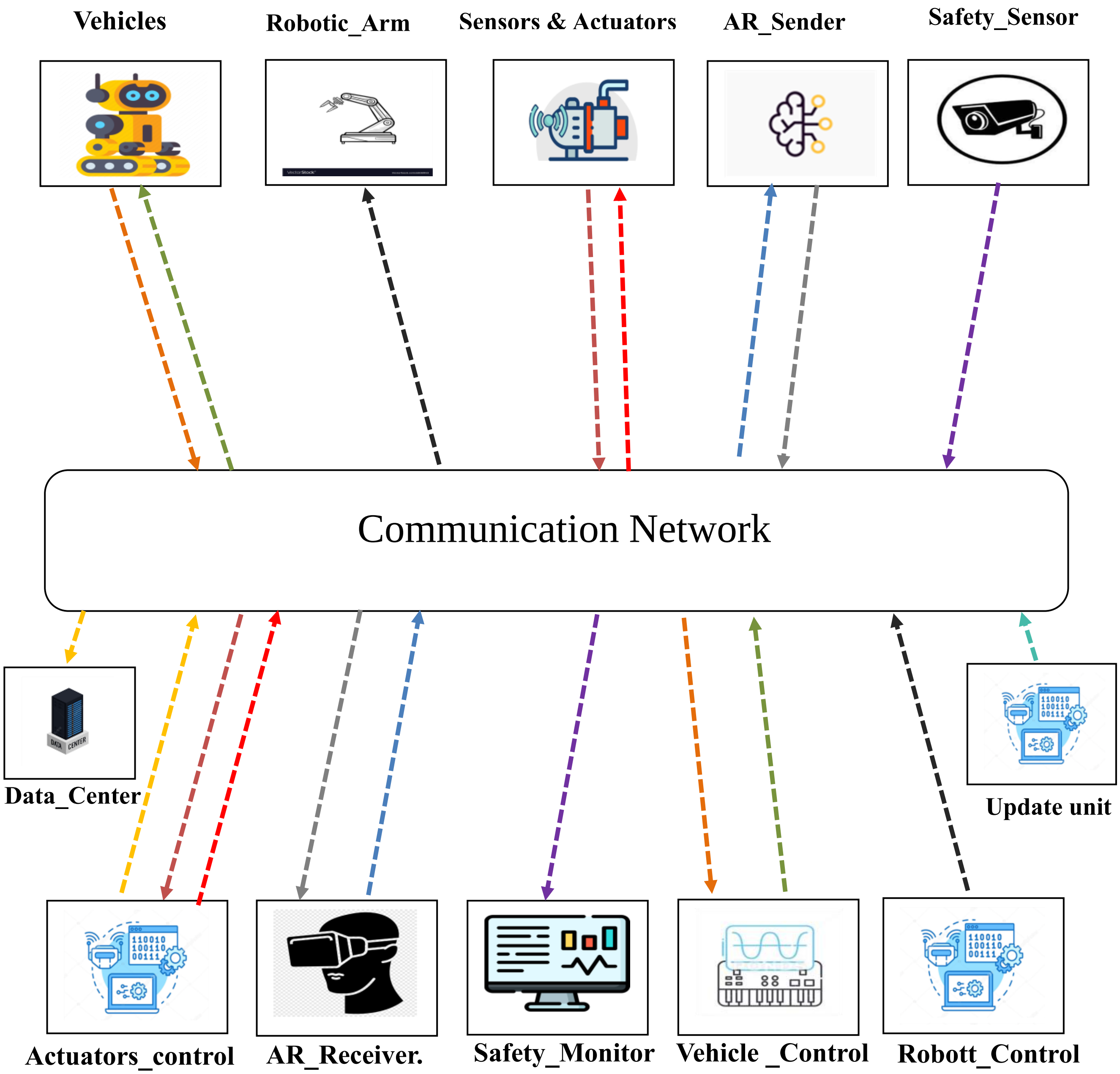

**Fig -4- Elements and messages flow of the adopted industrial model**

This model includes the following applications:

**Condition Sensors:** A collection of monitoring sensors linked to machinery and equipment to assess their operational conditions then transmit regular data to a local controller.

**Actuators Control:** local controller which receiving and analyzing the data from the condition monitoring sensors, it sends a control signal to carry out a specific task in the case that the conditions are defective.

**Safety Sensor:** Safety sensors have been supposed to monitor the conditions in the surrounding environments if any changes that may take place periodic data have been sent to the monitoring unit in order to alert it about any emergency situation that may occur. For example, if a fire or gas leak occurs, the appropriate decision must be taken regarding it.

**Safety Monitor:** Collect data on a regular basis from safety sensors to keep an eye on the various factors and situations that are surrounding the workers and factory floor.

**Augmented Reality (AR):** One of the most significant applications that are utilized in modern systems is augmented reality. It is an interactive application that gives the user the ability to interact virtually with industrial units. This is accomplished by attaching physical objects and components to the digital reality, displaying them, and interacting through technological devices such as tablets and laptops. Asking a remote expert to fix a problem by aiming a high-resolution camera at industrial machinery and transmitting video streams with some information to the expert to interact with them and provide the necessary notes to resolve the issue is one of the application's main uses, which has been assumed in this work. AR_Sender and AR_ Receiver are the elements which represented this application in the adopted model.

**RoboticArm and Control**: Industrial units have a variety of continuous production processes, and these production processes require effort to complete. Some examples of these production processes include the processes of packaging, sorting, and sealing the product. These processes require the presence of controlled automated arms that carry out them with high precision without any delay or error. This application which considered as remote-control application has been included into the model by use of two units, the controller for sending control signals and the arm in order to carry out these objectives.

**Vehicles and Control**: Mobile vehicles have been included into the model. These vehicles are meant to symbolize the application of automated guided vehicles application, which are utilized for the purpose of transporting items within the industrial unit. Vehicles and Vechiles_control in the model are represent this application. As the vehicles move inside a particular area, they transmit data to their control unit. The control unit then sends control signals to the vehicles in concern if there is a condition that requires changing the path that these vehicles take or modifying some of the functions that they do.

**Update unit:** This unit is responsible for the generation of best-effort data, which is then distributed to all applications in the network for the purpose of updating the software of these applications.

**Data_Center:** A data storage unit receives reports from the Actuators_control unit about the condition of the industrial unit after analyzing the received data from condition sensors.

### 4.2- The suggested communication network for the adopted model

The communication network model that has been designed for the typical industrial case has been shown in the Figure 5.

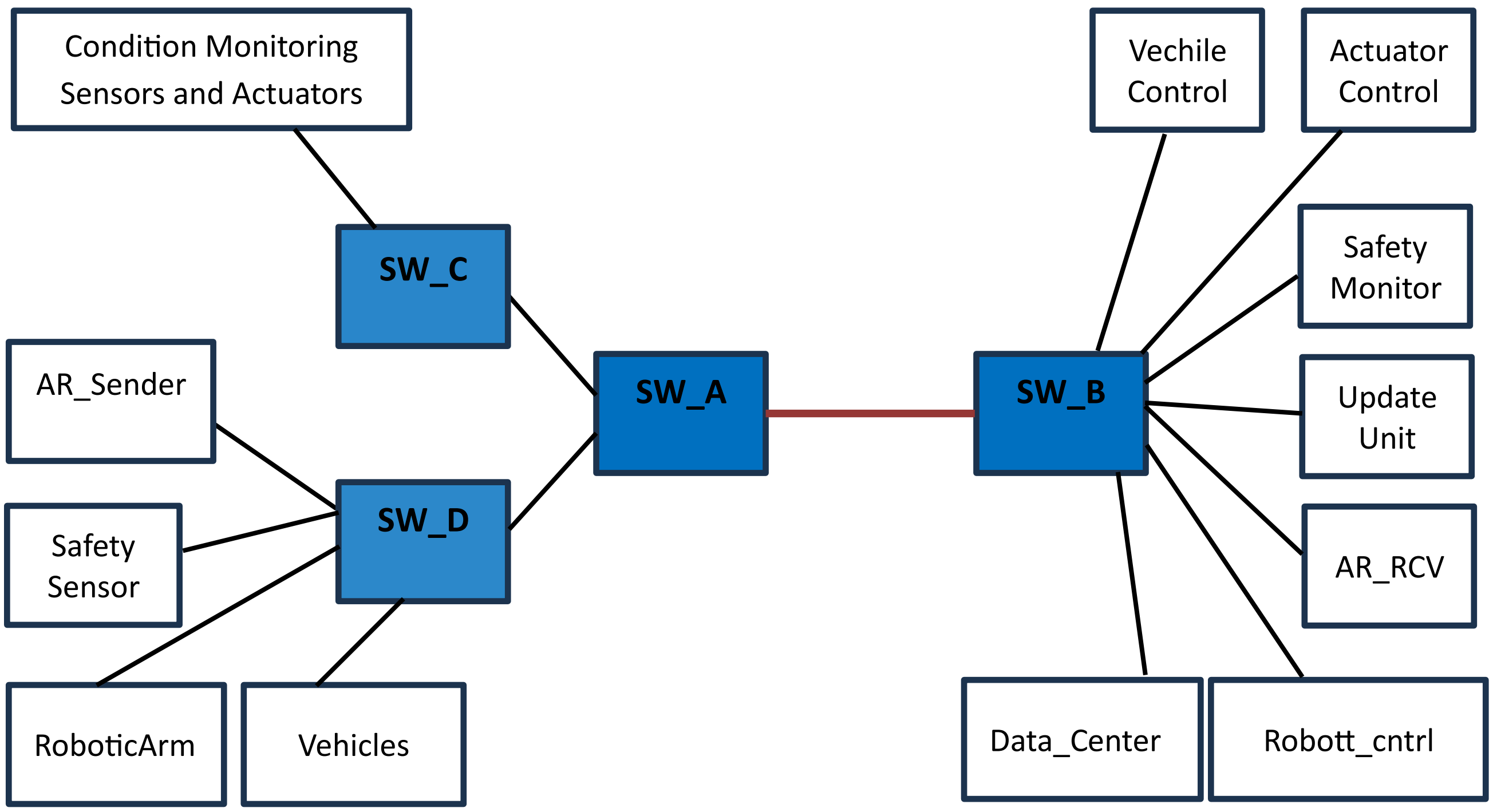

**Fig -5- Network Model of Industrial Case**

It is obvious that the suggested communication network depends on a wired network based on ethernet technology. This refers to provide a suitable shielding for industrial purposes in addition to handle the available commercial types of communication network technologies. It comprises a set of applications that represent a typical industrial case, interconnected by ethernet switches that support time-sensitive algorithms, linked with 100 Mbps Ethernet link that employs in abondance commercially for the industrial cases, except the core link between switches A and B, which has been selected of 1Gbps Ethernet link to capture the expected load in the backbone of communication network. Switches C and D in the network have been used to distribute and balance the load, enhancing the reliability of network applications. Switch A connects edge networks to the network backbone, while switch B finalizes the connection of all components to the designated network. Model specifications, number of nodes and the size of the data have been shown in the Table-4-.

**Table -4- Specifications of Model**

| Application | Representing Node | No. | Frame Size (B) | Frame Interarrival time |
|---|---|---|---|---|
| AR | AR_Sender | 1 | 1500 | 12µs |
| AR | AR_Receiver | 1 | 1500 | 40ms |
| Condition Monitoring | Condition_Sensor (CS) | 10 | 500 | 100ms |
| | Actuators | 10 | ----- | |
| Safety | Safety Sensor | 1 | 500 | 100ms |
| Automated Guided Vechile | Vechile | 2 | 1500 | 500µs |
| Remote Control | robottcontroller | 1 | 354 | 400µs |
| | Actuators Control | 1 | 354 | Depend on condition |
| | Vehicles Control | 1 | 354 | Depend on condition |

Network model has been simulated using OMNeT++ simulator, The main framework of OMNET++ is INET framework, which includes most of the components of wired and wireless networks, which can be reused in the designed models, In addition to this framework, NeSTing Framework[27] is another framework that specialized in simulating time-sensitive networks (TSN), which can be added to the work environment when designing the proposed models.

# 5- The Adopted Scenarios and Assumptions

## 5.1- Basic Scenario

A conventional Ethernet network scenario has been introduced as the basic scenario of the model without priorities by (First-In-First-Out (FIFO) scheduling method) to serve the network load by single queue at network switches. According to this method, there is no guarantee with respected to quality of service for the applications hence the data is served by the concept of first in – first out.

## 5.2- Time Sensitive Network Scenarios

Time-sensitive mechanisms (SP, CBS, TAS, FP) which explained previously have been simulated in these scenarios of the model. According to that, the applications of a network have been categorized into eight classes which are sent to eight separate queues in the network switches depend on the value in the 3-bit PCP field of the frame header which determined by the application's requirements, Consequently, the assigned PCP value of each application utilized in the model has been illustrated in Table-5-.

**Table -5- PCP value of applications**

| No. | Application | PCP Value |
|---|---|---|
| 1 | Remote Control | 7 (highest) |
| 2 | Safety and Protection | 6 |
| 3 | Augmented Reality (AR) | 5 |
| 4 | Automated Guided Vechile | 4 |
| 5 | Condition Monitoring | 3 |
| 6 | Update | 2(lowest) |

### 5.3 – Enhanced Scenario

In this scenario, edge computing concept as well as TSN methods have been exploited to processing the AR application data near to its source of generation, where the bit rate is large due to video nature of AR application which may effect on other applications in context of traffic received and delay. In this context, AR data has been compressed by properiate compression techniques based on edge-computing technology to assimilate the huge size of data to save available network resources. H.265 is the compression technique which has been used in this scenario to compress the AR data. There are several factors to measure the efficiency of the compression method, the compression ratio, a critical metric for assessing the efficacy of a compression technique, can be expressed in Equation 1. As well as the Peak Signal to Noise Ratio (PSNR) value, which measures the quality of the compressed data by comparing it with the original data, which can be calculated through Equation 2 measured in db.

$$CR= \frac{\textbf{Uncompressed Size}}{\textbf{Compressed Size}} \text{---------------------(1)}$$

$$PSNR = 10 * \log10((max^2)/MSE)\text{---------(2)}$$

max: maximum possible pixel value.

MSE: is the mean squared error between the pixel values of the uncompressed and compressed video versions.

In this scenario, compression ratio has been practically determined using the well-known Handbrake software[28] which applied to a Full HD video recorded in an industrial environment while, PSNR has been empirically measured using the MSU Video Quality Measurement Tool[29]. Subsequently, the Mean Opinion Score (MOS) has been computed based on the relationship depicted in Figure 6 illustrating the score of delivered service quality when this score is on a 1 to 5 scale, the scores are 1: unacceptable, 2: Poor, 3: Fair, 4: Good, 5: Excellent. Table 6 represents the specifications for the edge computing server.

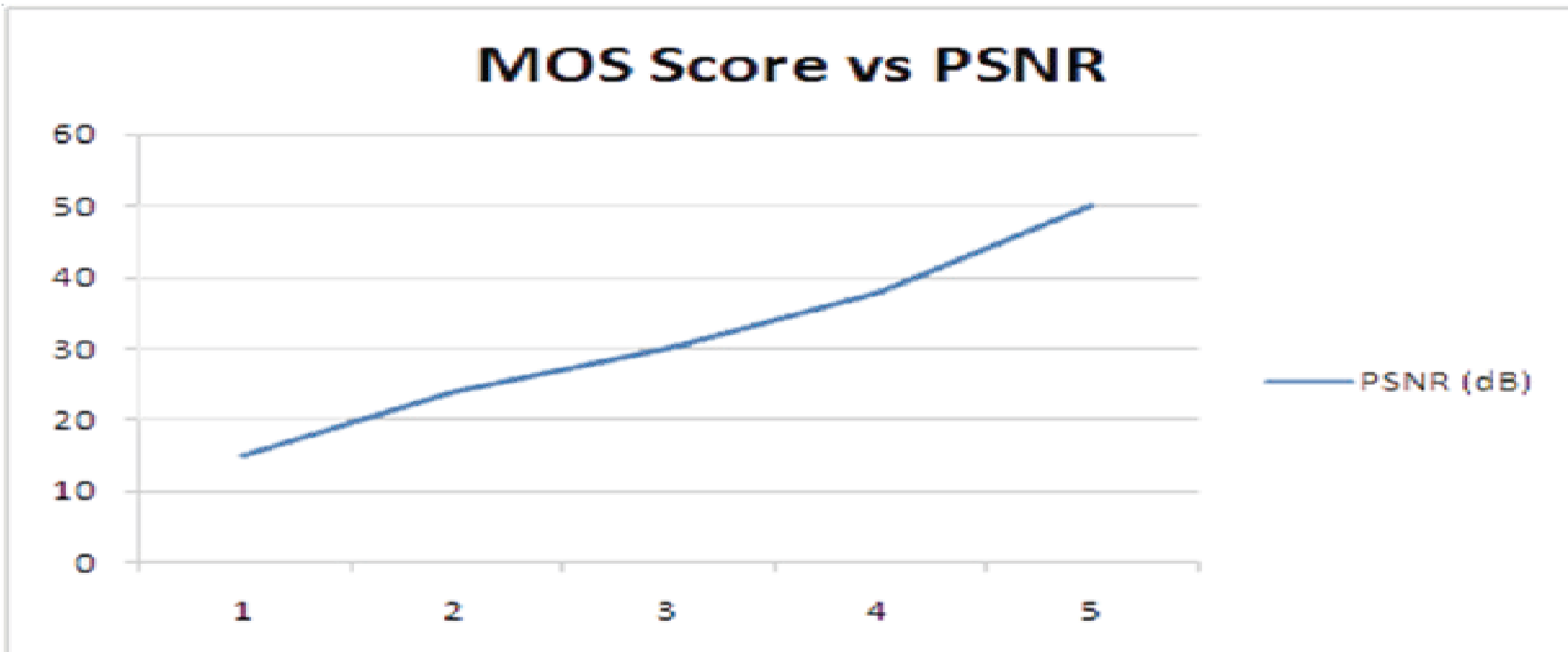

**Fig -6- MOS and PSNR Relationship**

**Table-6- Edge Server Specifications**

| NO. | Specification |
|---|---|
| 1 | CPU Intel Core i9 |
| 2 | 64 – bit OS (Operating System) |
| 3 | 32 GB RAM (Random Access Memory) |
| 4 | 500 GB dynamically allocated Storage |

## 5.4 – Upgrade Scenario in terms of physical cables

In the earlier scenarios, the bandwidth of the link that connected the terminal nodes to the network core was 100 Mbps. In this scenario, in addition to TSN mechanisms and emerging technologies, the bandwidth has been increased by replacing the 100 Mbps cables by 1 Gbps ethernet cables for all links of the network. This scenario investigates the effect of physical resources enhancement from cables point of view on the performance of the network. Figure 7 represents the flow chart of model methodology.

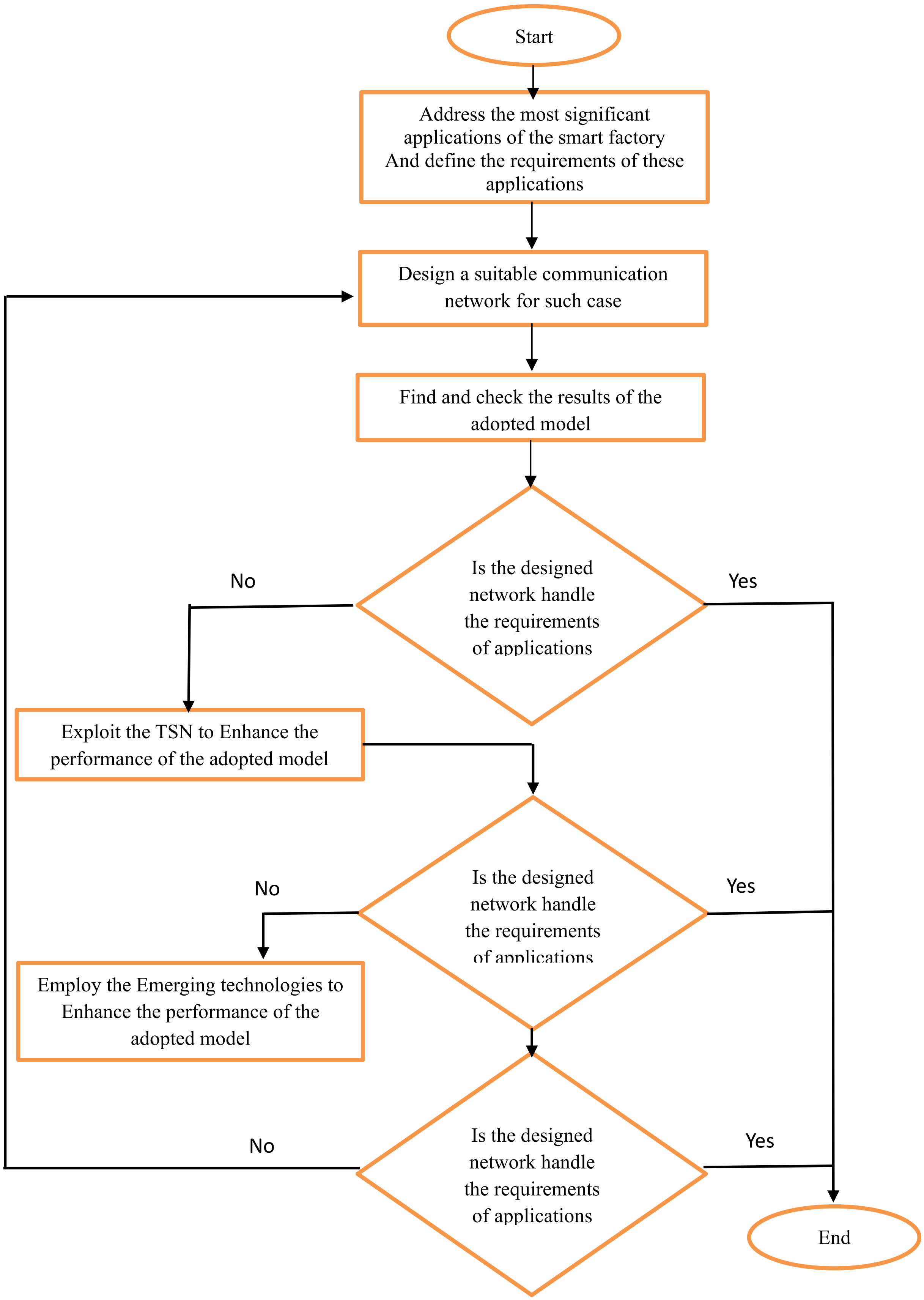

**Fig -7- Flow chart of Model Methodology**

## 6- Results and Discussion

This section examines and discusses the most notable outcomes of executing the proposed scenarios of model. The metrics of assumed applications of network model have been selected in context of traffic and delay time as explain in following:

- **Percentage of Received Data Reliability (RDR%)**: Which can be calculated through the following equation:

$$\frac{traffic\ received\ of\ application}{traffic\ sent\ of\ application} * 100$$

- **End to End delay:** The amount of time for frame transmission from source to destination.

The percentage of RDR value for every application in the basic scenario that follows the FIFO technique has been shown in Figure 8. Since the service is based on the arrival order of frames, this method does not ensure quality of service for certain applications. The figure shows that the majority of applications experience a decrease in the percentage of RDR traffic, indicating the algorithm's unreliability for these applications. This is particularly true for the highest priority robot controller application, which has a strong real-time component, since not all of its data reaches its destination. Due to the high generated bit rate—which reaches 995 Mbps—resulting from the use of a Full HD camera with a resolution of 1920*1080, the R/S value was the lowest for AR application approximately equal to 8%.

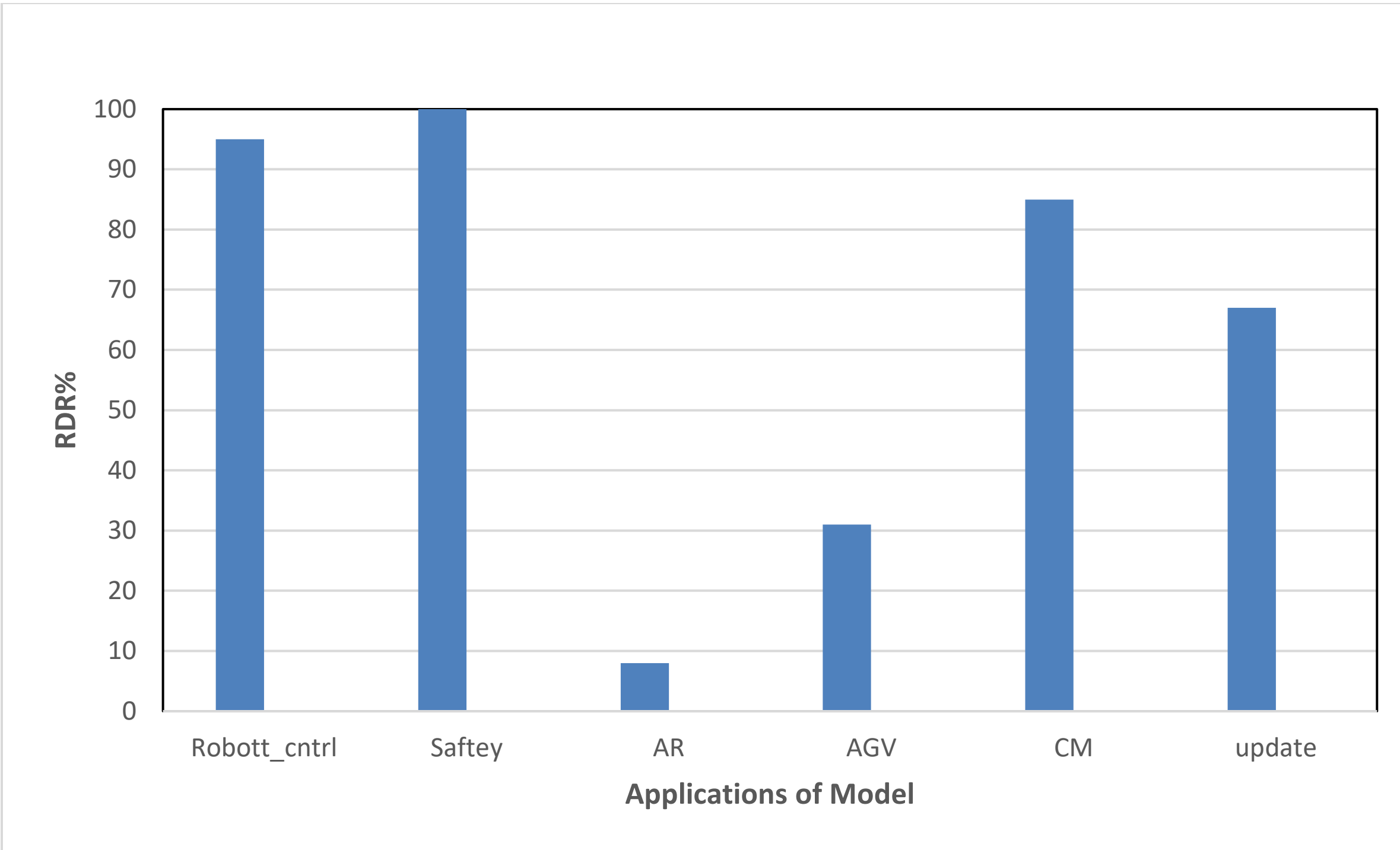

**Fig -8- RDR% for Model Applications in FIFO Scenario**

Figure 9 illustrates the end-to-end delay time, indicating a discrepancy between the delay times for applications and their respective requirements. The delay time for the remote-control application reaches 4 ms, whereas the requirement is between (0.1 and 1) ms. Similarly, the safety application exceeds 12 ms, surpassing the requirement of 10 ms, the permissible maximum limit for the application**.** The delay time for the AR application was markedly greater than other apps which approximately equal to 900ms due to longer waiting for the application frames in the queue, resulting in the loss of the majority of them.

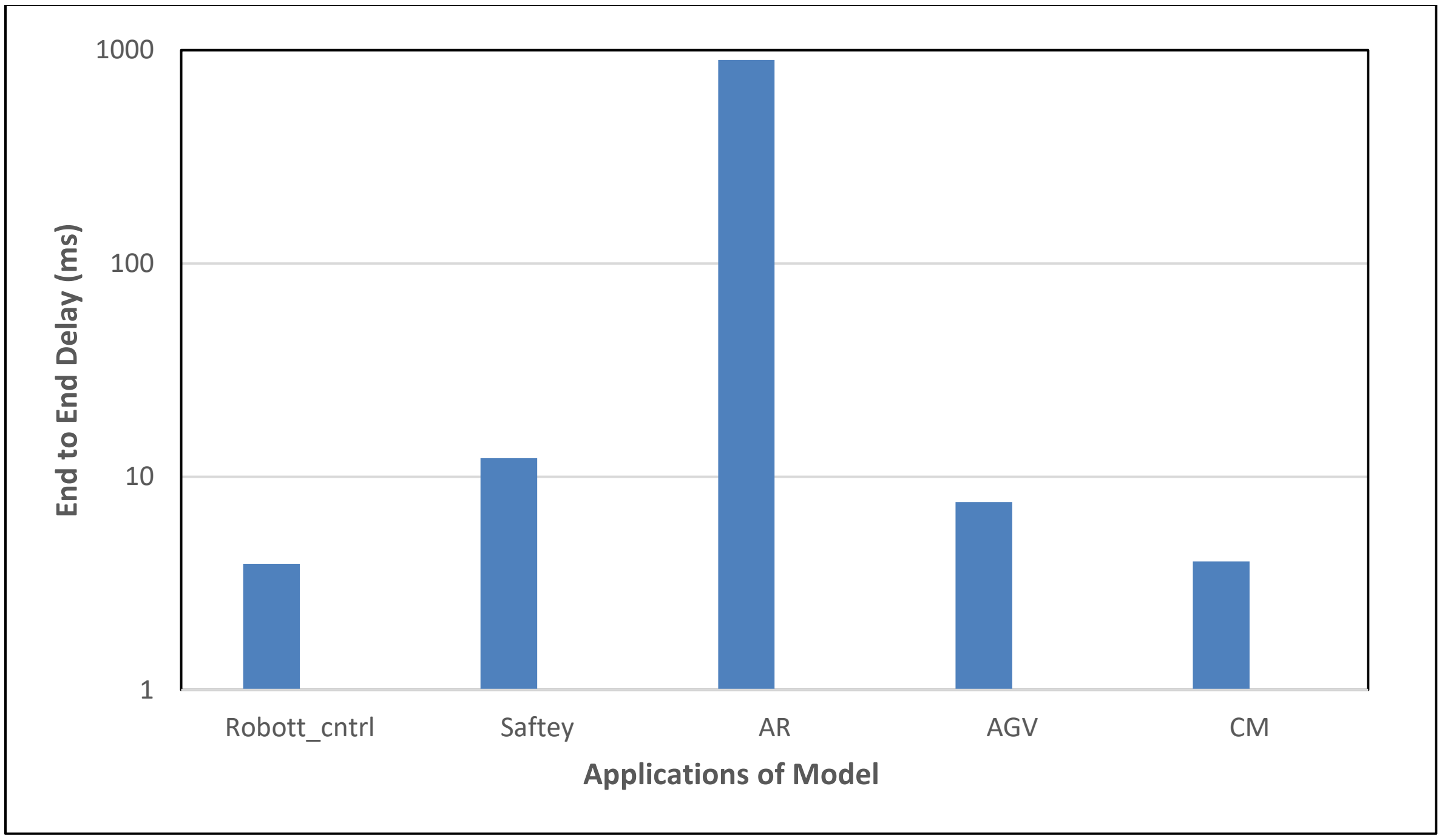

**Fig -9- End to End Delay for Model Applications in FIFO Scenario**

The results for TSN scenario are shown in Table -7-, where applications have been categorized into several classes and the highest priority applications have been served firstly, then the rest. The SP method has been considered as the default method for serving data from queues, where highest priority applications are always given the service. As seen in the table, the Remote Control and Safety applications have been completely serviced. Conversely, it is observed that the AR application utilizes the full bandwidth of the link to transmit just 10% of its data, approximately 97 Mbps, thereby entirely restricting service to other less prioritized applications, as indicated in the results which showing that no data received for AVG and CM applications. CBS is the second method for serving applications from queues, however it is less stringent than SP because it enables the distribution of service among queues based on the Idle Slope value. In general, it is utilized for video applications, So, it has been allocated as a default to send AR data from its queue (5). The results have been collected at Idle Slope values ranging from 0.1 to 0.9. At a value range of 0.1 to 0.5, the data receiving percentage for the AR application diminished to 5%, which is near to the previous 10%, indicating failure in both cases. Conversely, the percentage for other applications (AVG, CM) increased to 100%. Following a value of 0.5, the condition getting worse, nearing the SP case at a value of 0.9 for Idle Slope; thus, the optimal value for Idle Slope, to be established in remaining scenarios is 0.5. In the TAS scenario, the protected window has been set to 140 µs, which is the approximate arrival time of remote-control application frames from source to destination. The unprotected window is 220 µs and the guard band to 40 µs. The combined three periods equal to 400 µs, which is the interval time between transmission remote control application frames, it is seen that by controlling queue gates, this approach favors time-sensitive applications over other applications, resulting in a lower RDR ratio for the AR and AVG apps than in the CBS (0.5) scenario. Conversely, in the FP scenario, the results had been better as the queue gates remained open; only the queues of lowest priority have been interrupted to transmit frames from the highest priority queue.

**Table – 7- RDR% for Model Applications in TSN Scenarios**

| **Application** | **TSN Mechanisms** | | | | | | | |
|---|---|---|---|---|---|---|---|---|
| | **SP** | **CBS (0.1)** | **CBS (0.3)** | **CBS (0.5)** | **CBS (0.7)** | **CBS (0.9)** | **TAS** | **FP** |
| **Robottcntrl** | 100 | 100 | 100 | 100 | 100 | 100 | 100 | 100 |
| **Safety** | 100 | 100 | 100 | 100 | 100 | 100 | 100 | 100 |
| **AR** | 10 | 1 | 3 | 5 | 6 | 8 | 2 | 5 |
| **AGV** | 0 | 100 | 100 | 100 | 72 | 25 | 13 | 100 |
| **CM** | 0 | 100 | 100 | 100 | 1 | 0 | 100 | 100 |
| **update** | 70 | 70 | 70 | 70 | 70 | 70 | 35 | 70 |

Table -8- illustrates the end-to-end delay for TSN scenarios. The results indicate that the delay time for the remote-control application is minimal across all scenarios, particularly in the TAS scenario, where its equal to 0.13 ms, and is somewhat more in the FP scenario due to his high priority. Because the Safety application only requires a small bandwidth of roughly 1 Mbps, it did not experience loss or delay in any scenario, and its delay time has slightly higher than remote control application as well as its value is constant across all scenarios. The AR application, on the other hand, has the worst delay of all TSN scenarios because of the size of its data and the network's inability to transmit it. As a result, the majority of the data is lost, and the rest have been stayed in the queue for a long time before being served. The delay for the remaining applications varies depending on the scenario. In conclusion, it is generally preferable to use the SP algorithm for all queues in TSN scenarios, with the exception of queue 5, which is assigned to the AR application and should be based on the CBS algorithm with IdleSlope of 0.5. The FB mechanism should be used in place of TAS, which has negative effects on all other applications except the critical ones. However, the AR application continues to suffering failures and requires alternative solutions, as seen in the following scenarios.

**Table -8- End to End Delay (ms) for Model Applications in TSN Scenarios**

| **Application** | **TSN Mechanisms** | | | | | | | |
|---|---|---|---|---|---|---|---|---|
| | **SP** | **CBS (0.1)** | **CBS (0.3)** | **CBS (0.5)** | **CBS (0.7)** | **CBS (0.9)** | **TAS** | **FP** |
| **Robottcntrl** | 0.3 | 0.3 | 0.3 | 0.3 | 0.3 | 0.3 | 0.13 | 0.15 |
| **Safety** | 0.37 | 0.37 | 0.37 | 0.37 | 0.37 | 0.37 | 0.37 | 0.37 |
| **AR** | 902 | 905 | 903 | 902 | 902 | 902 | 903 | 902 |
| **AGV** | 0 | 0.76 | 0.8 | 0.9 | 11 | 30 | 58 | 0.9 |
| **CM** | 0 | 1.2 | 2.2 | 7.4 | 13.5 | 0 | 2.3 | 8 |

The RDR% results for applications when the network is developed to 1 Gbps for the peripheral links have been displayed in Figure 10. It can be seen that the AR application's RDR percentage has been increased to 100. However, it should be highlighted that other, lower priority applications do not have any RDR percentage. This is because the AR application reserved the entire bandwidth to transfer its data at a rate of around 1 Gbps, which caused no service to be provided to these applications.

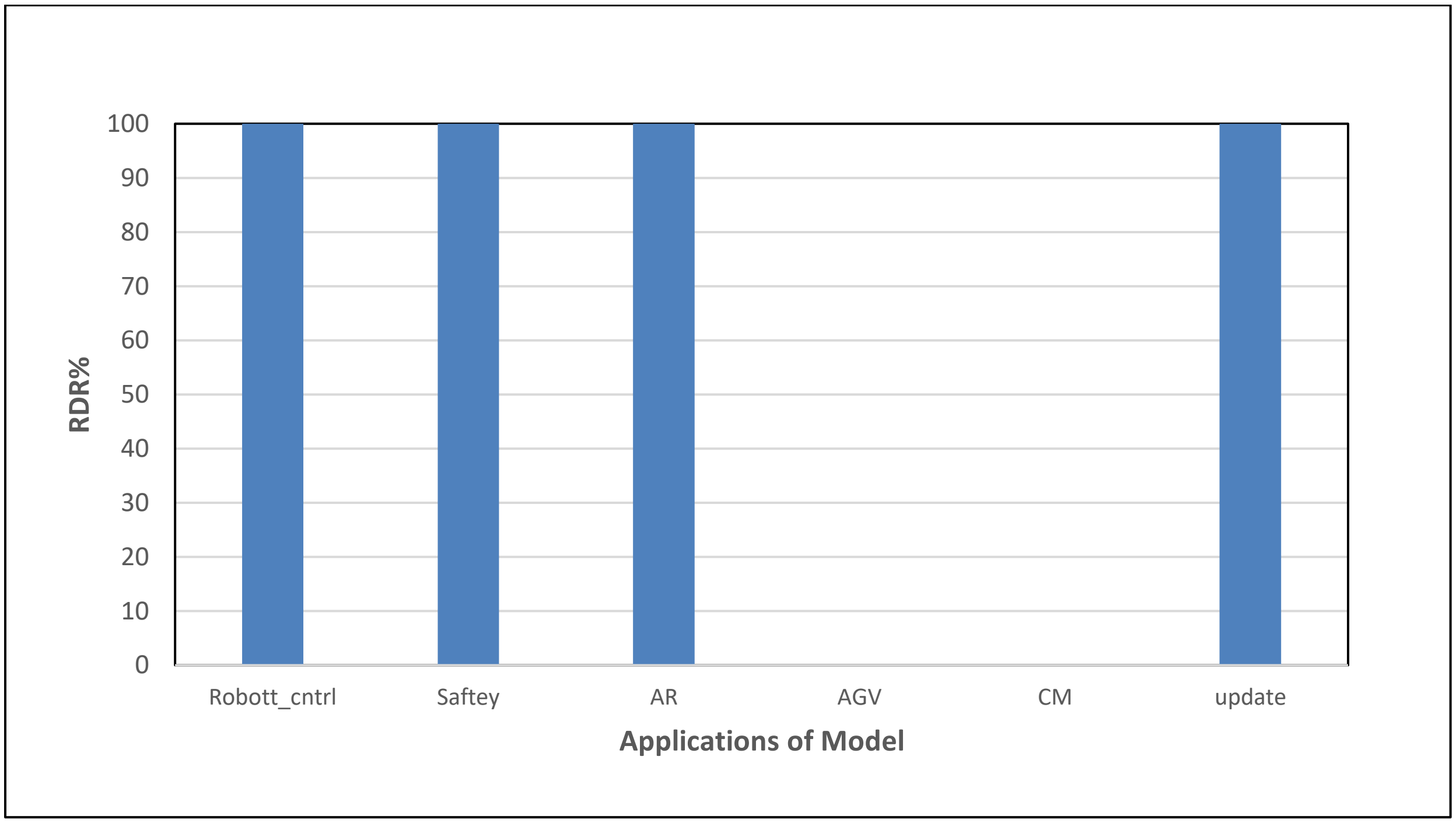

**Fig -10- RDR% for Model Applications in Upgrade Scenario**

On the other hand, network upgrades have resulted in a significant decrease in the delay time for data-received of applications, as noted in Figure 11, the delay time for the remote-control application decreasing from approximately 0.15 ms at 100 Mbps to 40 µs and for safety application from 0.37 ms to 60 µs, while the delay time remained the highest in the AR application, approximately to 430 µs. However, since the AVG and CM applications have not received any service, no value has been recorded for the delay time.

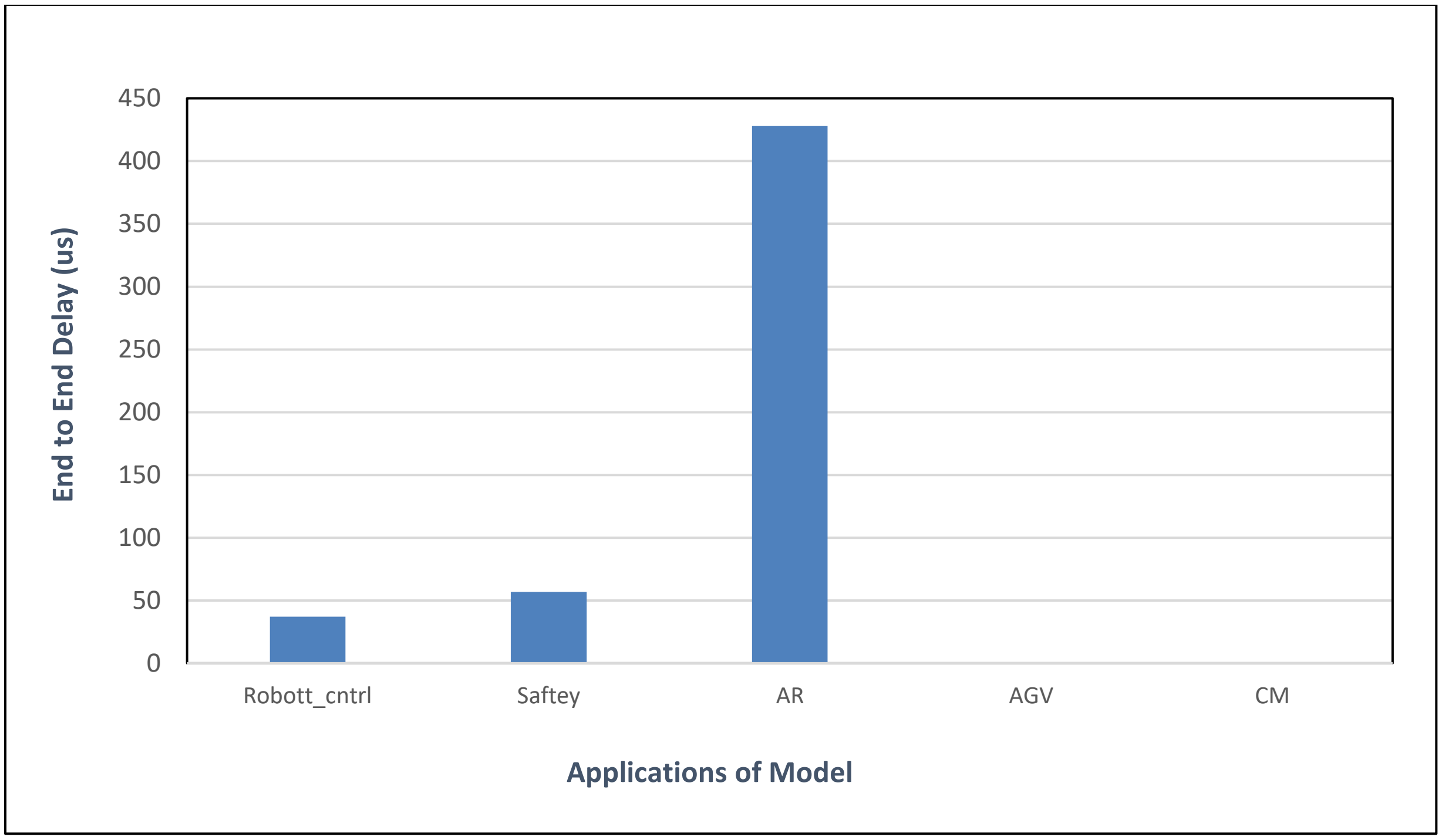

**Fig -11- End to End Delay for Model Applications in Upgrade Scenario**

As it can be seen from previous scenarios, AR application can be considered the main reason why the network isn't reliable enough to serve all applications, which exhausts network resources due to the huge size of its data therefore, AR data has been compressed with several compression ratio. The optimal compression ratio for achieving reliability is 22:1 with a bandwidth of 100 Mbps for peripheral links, however at 1 Gbps, less compression ratio of 2:1 is required as seen in Figure 12. To ensure that the level of AR data compression has not been affected on application quality, the PSNR value has been empirically measured which it equal to 36 dB for 22:1 compression ratio that is deemed acceptable according to the relationship illustrated in the Figure 6. Figure 13 (a&b) shows a sample of the recorded video before and after the compression process. In addition to using edge computing in this scenario, the CBS with IdleSlope of 0.5 has been chosen to serve AR data from queue 5, while SP has been selected to serve data from other queues. Time-sensitive applications are supported by the FP mechanism.

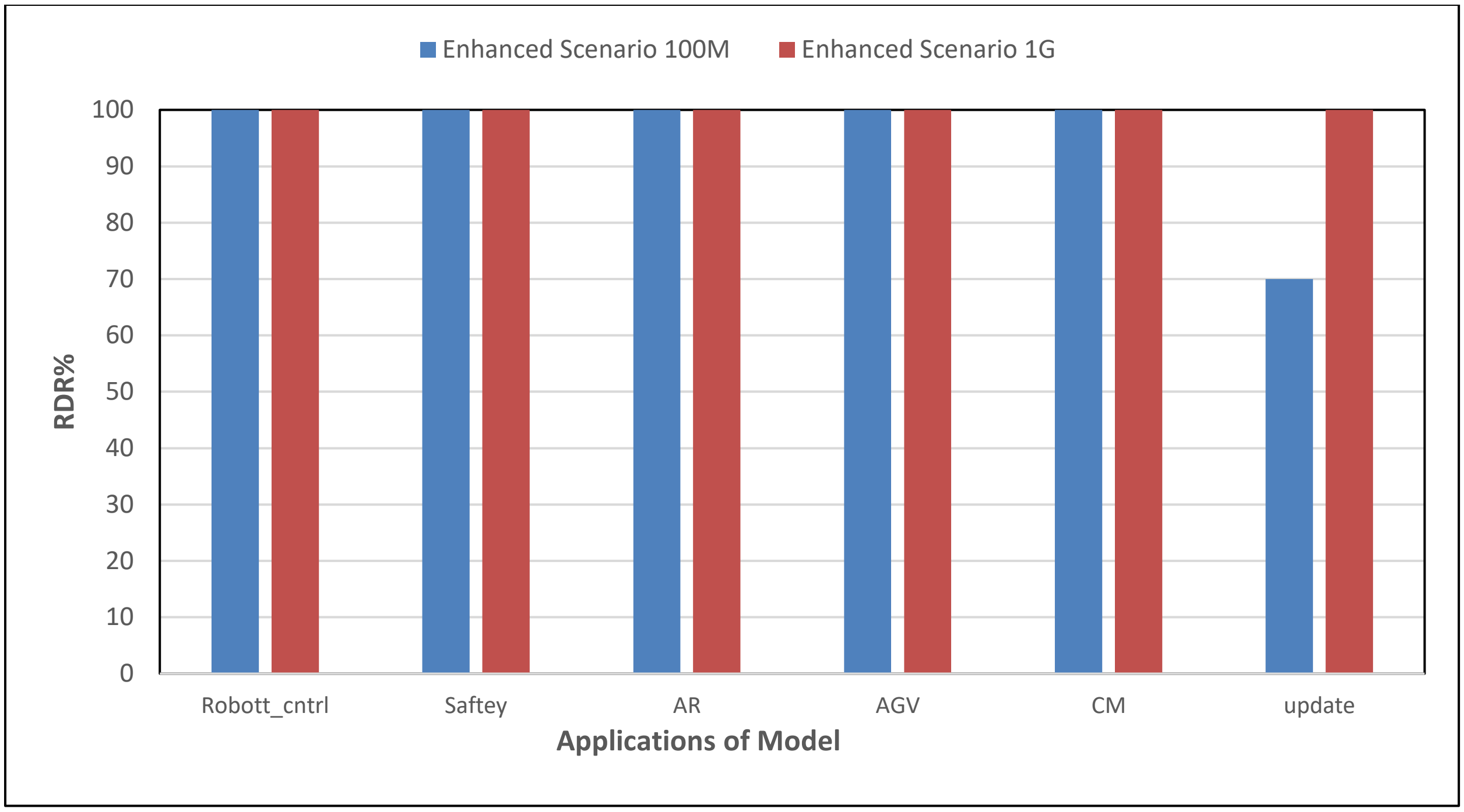

**Fig -12- RDR% for Model Applications in Enhanced Scenario**

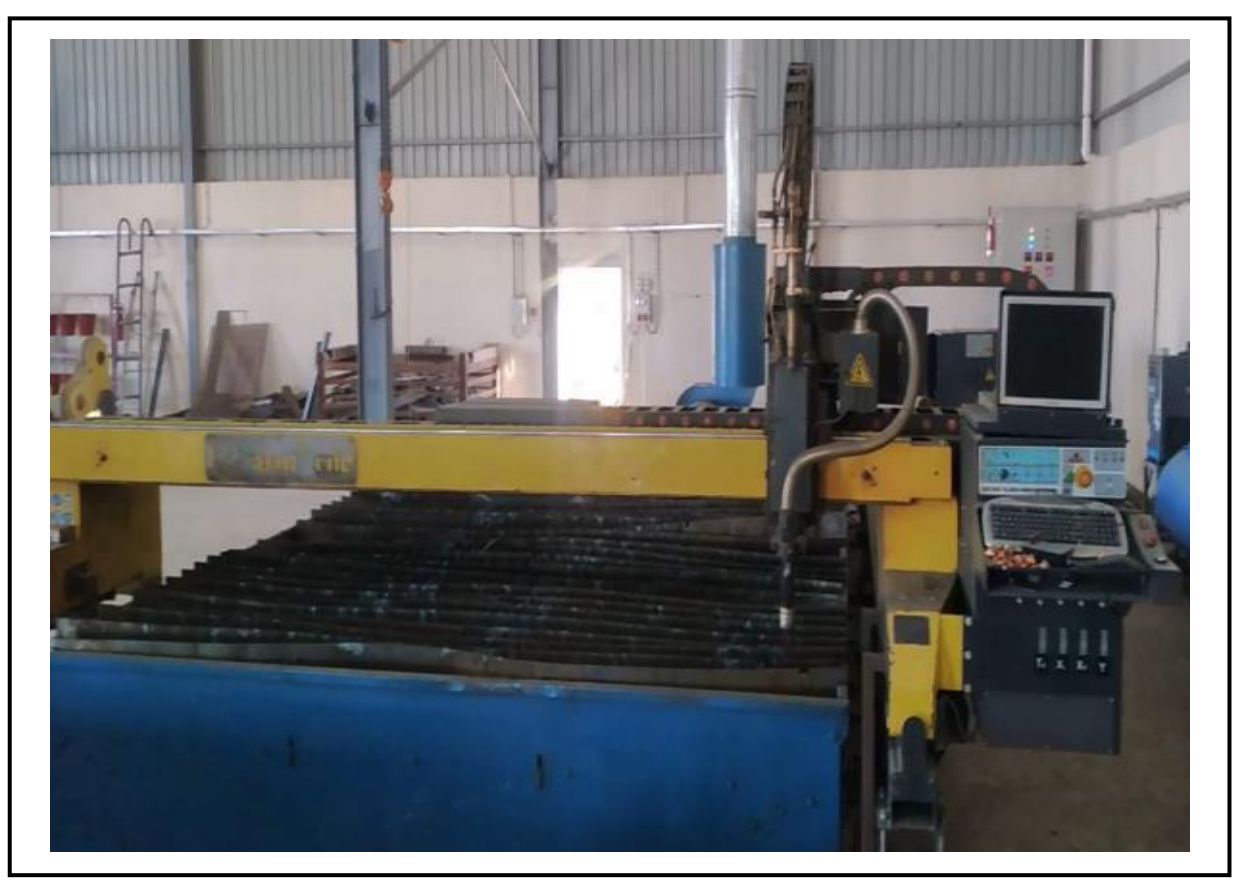

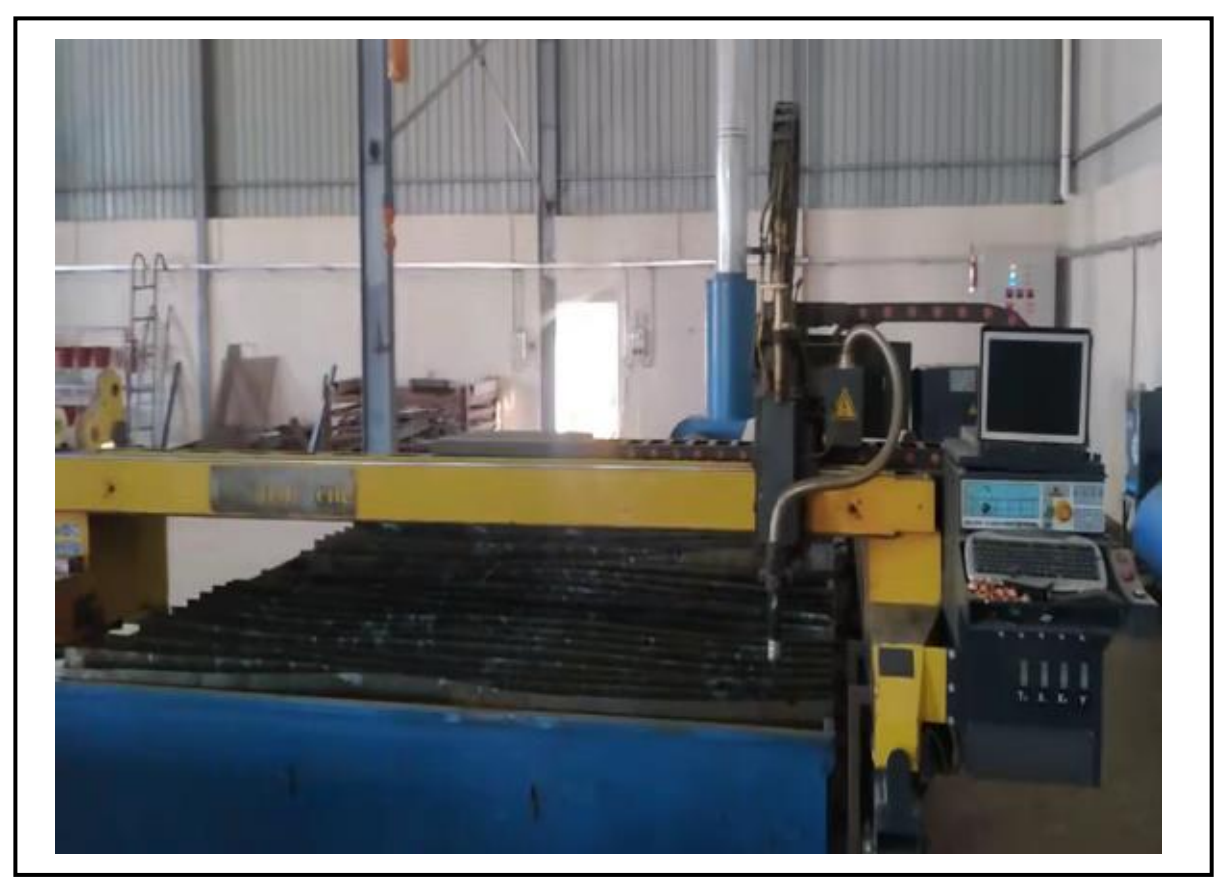

**a** **b**

**Fig -13- AR sample before and after compression**

The delay time for the model applications at 100 Mbps and 1 Gbps has been shown in Figure 14. It is evident that all applications have delay values within the requirements of applications. On the other hand, The AR application has the largest delay value due the additional processing time which required to process its data at the edge server. This is among the drawbacks of data compression base on edge computing concept.

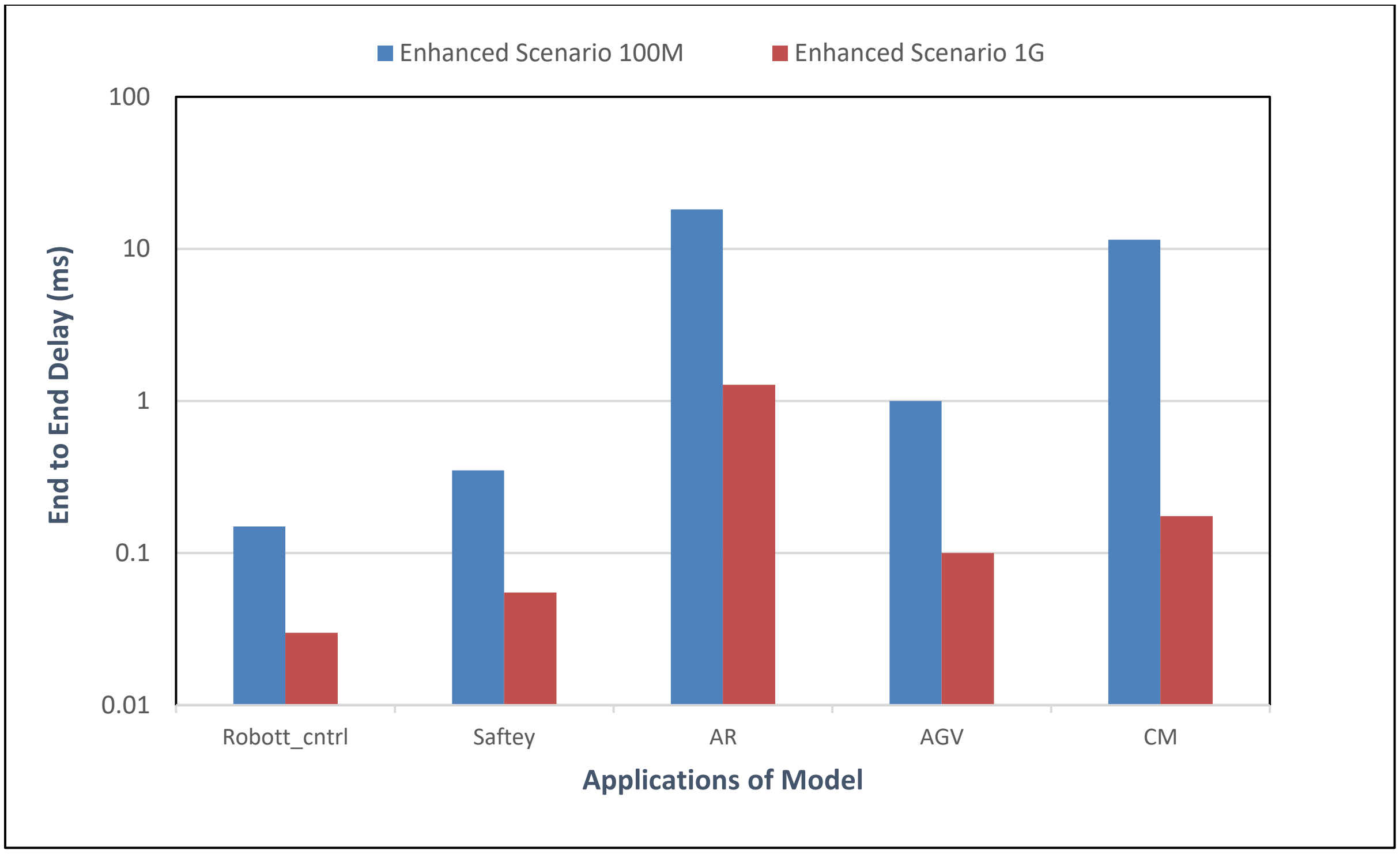

**Fig -14- End to End Delay for Model Applications in Enhanced Scenario**

Figure 15 illustrates the delay time for the control signals which produced by the local controllers (Actuators_control and Vehicles_control) when a specific circumstance arises requiring an instance response. It can be seen that, except the FIFO and TAS scenarios, the delay time for these signals upon transmission within the specified limits of the requirements of these applications**.** Conversely, it has been observed that the remaining scenarios have not been registered any values for these signals because their data has not been reached the controllers due to failure of transmission by the network.

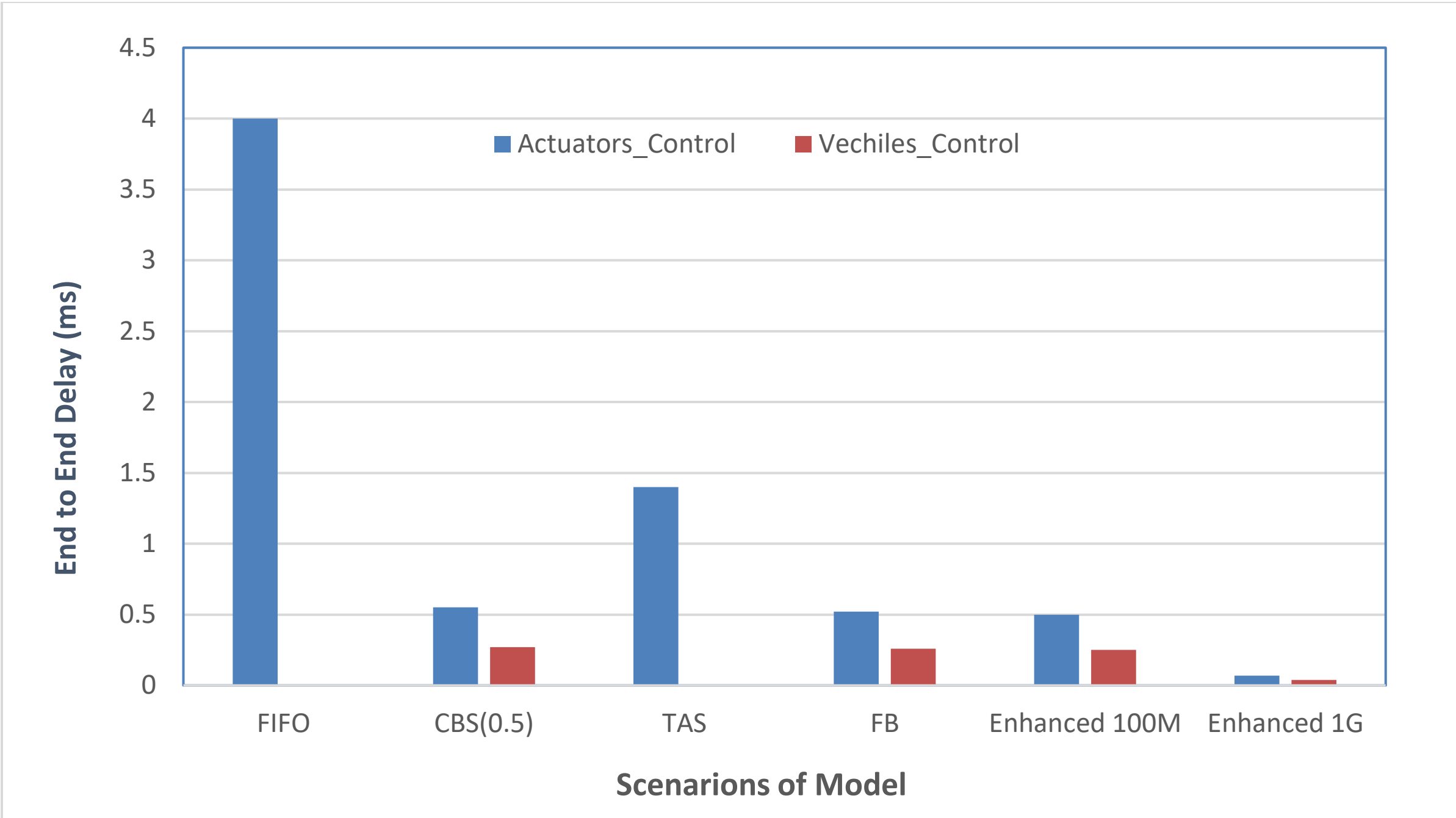

**Fig -15- End to End Delay for Control Applications of different scenarios**

A comparison of all realized scenarios according to how well they meet the requirements of the main smart factory applications which selected in this work has been shown in Table -9-.

**Table-9- Handling the applications requirements according to network scenarios**

| Scenario | The Main of Smart Factory Applications | | | | | | | | | |
|---|---|---|---|---|---|---|---|---|---|---|
| | Remote_Cntrl | | Safety | | AR | | AVG | | CS | |
| | RDR | Delay | RDR | Delay | RDR | Delay | RDR | Delay | RDR | Delay |
| **Basic** | × | × | √ | × | × | × | × | √ | × | √ |
| **TSN(SP)** | √ | √ | √ | √ | × | × | × | × | × | × |
| **TSN(CBS)** | √ | √ | √ | √ | × | × | √ | √ | √ | √ |
| **TSN(TAS)** | √ | √ | √ | √ | × | × | × | × | √ | √ |
| **TSN(FP)** | √ | √ | √ | √ | × | × | √ | √ | √ | √ |
| **Upgrade** | √ | √ | √ | √ | √ | √ | × | × | × | × |
| **Enhance** | √ | √ | √ | √ | √ | √ | √ | √ | √ | √ |

## 7- Conclusion

This research presents a robust communication framework for next-generation industrial networks, specifically targeting the requirements of smart factories in the context of Industry 4.0. By integrating Time-Sensitive Networking (TSN) with edge computing, the proposed model effectively addresses the challenges posed by diverse applications requiring varying levels of latency, bandwidth, and reliability. The experimental results demonstrate that the enhanced communication network achieves impressive performance metrics, including a latency of less than 1 millisecond for real-time applications and a packet delivery success rate of 99.999%. These outcomes highlight the feasibility of deploying TSN mechanisms, such as Strict Priority (SP) and Frame Preemption (FP) in conjunction with advanced data compression techniques like H.265, to optimize network performance. Moreover, this work emphasizes the critical role of edge computing in managing heavy data loads, ensuring that applications such as augmented reality can function seamlessly without degrading the performance of real-time systems. The balance achieved between various application requirements underscores the potential for this integrated approach to support complex industrial environments. Future research may focus on further refining the model to incorporate adaptive mechanisms that respond dynamically to changing network conditions and application demands, thereby enhancing the resilience and efficiency of smart factory operations. Overall, the findings of this work contribute significantly to the ongoing evolution of industrial communication networks, paving the way for more intelligent and responsive manufacturing systems.

## Acknowledgment

The authors express their gratitude to the University of Mosul, College of Engineering, Department of Electrical, for providing the necessary facilities that contributed to enhancing the quality of this paper